\documentclass[12pt]{article}

\usepackage{amssymb}

\usepackage{amsthm}

\newcommand{\refbr}[1]{(\ref{#1})}
\newcommand{\beq}{\begin{equation}}
\newcommand{\eeq}{\end{equation}}
\def\goesto{\mathop{\rightarrow}}

\usepackage{t1enc}

\begin{document}

\title{Amdahl's and Gustafson-Barsis laws revisited}

\author{Andrzej Karbowski\\
Institute of Control and Computation Engineering,\\
        Warsaw University of Technology,\\
        ul. Nowowiejska 15/19, 00-665 Warsaw, Poland,\\
NASK (Research and Academic Computer Network),\\
 ul.~W\k{a}wozowa 18, 02-796~Warsaw, Poland\\
 E-mail:~A.Karbowski@ia.pw.edu.pl}

\maketitle
\begin{abstract}
The paper presents a simple derivation of the Gustafson-Barsis law from the Amdahl's law.
In the computer literature these two laws describing the
speedup limits of parallel applications are derived separately.
It is shown, that treating the time of the execution of the sequential part
of the application as a constant, in few lines the Gustafson-Barsis law
can be obtained from the Amdahl's law and that the popular claim, that Gustafson-Barsis law overthrows
Amdahl's law is a mistake.
\end{abstract}
{\bf Keywords:}
parallel computing, distributed computing, speedup

\date{ }
\parindent 2em

\thispagestyle{empty}
\section{Introduction}

The Amdahl's law formulated about four decades ago \cite{amdahl} is considered to be one of
the most influential concepts in parallel and distributed processing \cite{top}.
It describes the limits on the speedup obtained owing to the execution of
the application on the parallel machine with relation to the single-processor,
sequential machine. More precisely, Amdahl's law says, that the speedup of an
application obtained owing to the execution on the parallel machine cannot be
greater that the reciprocal of the sequential fraction of the program.
Speedup restrictions resulting from Amdahl's law prevented
designers from exploiting parallelism for many years, being a nuisance to vendors of parallel
computers \cite{lewis-rew}.
The rescue came from Sandia Labs. On the basis of some experiments,
Gustafson \cite{gustafson} claimed that "the assumptions underlying Amdahl's 1967 argument
are inappropriate for the current approach to massive ensemble parallelism".
Furthermore, Gustafson formulated "an alternative to Amdahl's law suggested by E. Barsis at Sandia". The so-called Gustafson-Barsis law is said to vindicate the use of massively parallel processing \cite{reilly}, \cite{enc.CS}.
However, in the author's opinion, when we analyze deeper both laws, we will see,
that Gustafson's results do not refute the Amdahl's law, and the Gustafson-Barsis law can be directly derived from the Amdahl's law.

\section{Amdahl's and Gustafson-Barsis laws in the original form}

Although in the original Amdahl's paper \cite{amdahl} there were no equations, basing on
the verbal description
one may present his concept formally. The way  of our presentation
is similar to that of \cite{lewis}, \cite{lewis-rew}, with only one difference, which will be explained later on.
It is assumed in the model, that the program consists of two parts: sequential and parallel. While
the time of the execution of the sequential part for a given size $n$ is the same on
all machines, independently of the number of processors $p$, the
parallel part is perfectly scalable, that is,  the time of its
execution on a machine with $p$ processors is one $p$-th of
the time of the execution on the machine with one processor. Let us
denote by $\beta(n,p)$ the sequential fraction of the total real-time $T(n,p)$
of the execution of the program on a machine with $p$
processors (the mentioned difference introduced here is treating both the fraction $\beta$ and time $T$ as functions
of $n$ and  $p$; it will prove to be very useful afterwards).

With this notation we may calculate
the sequential part time $T_s$ for the given problem size $n$ from the expression
\beq
T_s(n)= \beta(n,1) \cdot T(n,1)
\eeq
and the parallel part time $T_p$, which is dependent on the problem size $n$ and the number of processors $p$, from the expression
\beq
T_p(n,p) = \frac{(1-\beta(n,1)) \cdot T(n,1)}{p}
\eeq
If we ignore communication costs and overhead costs associated with operating system functions, such as process creation, memory management, etc. \cite{lewis-rew}, the total time $T(n,p)$ will be the sum of sequential and parallel part time, that is
\[
T(n,p)=T_s(n)+T_p(n,p)= \beta(n,1) \cdot T(n,1) + \frac{(1-\beta(n,1)) \cdot T(n,1)}{p} =
\]
\beq
=\left[ \beta(n,1)  + \frac{1-\beta(n,1)}{p}\right] \cdot T(n,1)
\label{total.amd}
\eeq
From \refbr{total.amd} we get directly the formula for the speedup $S(n,p)$ obtained
due to the parallelization of the application:
\beq
S(n,p) = \frac{T(n,1)}{T(n,p)} = \frac{1}{\beta(n,1) + \frac{1 -
\beta(n,1)}{p}}
\label{pr.Amd}
\eeq
The formula  \refbr{pr.Amd} is called Amdahl's law.
It is seen, that in the limit
\beq
S(n,p) \goesto_{p \goesto \infty} \frac{1}{\beta(n,1)}
\eeq
It means, that even when we use infinitely many parallel processors,
we cannot accelerate the calculations more than the reciprocal
of the sequential fraction of the  execution time of the program on a sequential
machine. That is, for example, when this factor equals $\frac{1}{2}$,
the program can be accelerated at most twice, when $\frac{1}{10}$
-- ten times! Speedup restrictions resulting from Amdahl's law prevented
designers from exploiting parallelism for many years, being a problem to vendors of parallel
computers \cite{lewis-rew}.

The help came from Sandia Labs. In some experiments described by Gustafson \cite{gustafson} it was taken, that the run time was constant, while the problem
size scaled with the number of processors. More precisely, the time of the sequential part was independent, while the work to be done in parallel varied linearly with the number of processors. Since the time of the execution in Gustafson's paper \cite{gustafson} was normalized to 1, that is
\beq
T_s(n) + T_p(n,p) = 1
\label{sum.T}
\eeq
we had actually the equivalence
\beq
\beta(n,p) \equiv T_s(n)
\label{equi.beta.T}
\eeq
and
\beq
T_p(n,p) = 1 - \beta(n,p)
\eeq
Following Gustafson, a serial processor would require time $T_s(n)+T_p(n,p) \cdot p$ to perform the task, so the scaled speedup on the parallel system was equal:
\beq
S(n,p) = \frac{T_s(n)+T_p(n,p)\cdot p}{T_s(n)+T_p(n,p)} = T_s(n)+T_p(n,p)\cdot p
= p+(1-p)\cdot T_s(n)
\label{scal.speed}
\eeq
Using the equivalence \refbr{equi.beta.T} we may write \refbr{scal.speed} in the following form:
\beq
S(n,p) = p+(1-p)\cdot T_s(n)  = p+(1-p)\cdot \beta(n,p)
= p - (p-1) \cdot \beta(n,p)
\label{GustBar.oryg}
\eeq
The last equation is called Gustafson-Barsis law.

\section{The main results}

In the Gustafson's paper \cite{gustafson}, three things raise some doubts:
\begin{enumerate}
\item Mixing the problem size and the number of processors, treating both as
tightly connected ("the problem size scales with the number of processors")
\item Normalizing the time of calculations on the sequential machine
to 1 (eq. \refbr{sum.T}) for all problem sizes and numbers of processors
\item Treating assessment \refbr{GustBar.oryg} as a better alternative
to Amdahl's law, derived  independently, basing on different assumptions
\end{enumerate}
The truth is, that Gustafson-Barsis law is nothing but a different form
of Amdahl's law, and that better values of the speedup in the Gustafson's experiments with the growing size of the problem could be obtained directly
from the Amdahl's law.

To show this it is sufficient to notice, that for a given problem size $n$ there is a constant in all executions of the program, on machines with different
number of processors. This constant is the time of the execution of the sequential
part $T_s(n)$ for the given problem size $n$. It is independent of the number of processors $p$, that is:
\beq
T_s(n)=\beta(n,p) \cdot T(n,p) = const.,\;\;\;\; p=1,2,3,\ldots
\label{czas.sekw}
\eeq 
So, it will be for any $p=1,2,3,\ldots$
\beq
\beta(n,1) \cdot T(n,1) = \beta(n,p) \cdot T(n,p)
\label{rown.czasu.1p}
\eeq
From the equation \refbr{rown.czasu.1p} we get:
\beq
\beta(n,1)=\beta(n,p) \cdot \frac{T(n,p)}{T(n,1)}
\label{beta.1.zal}
\eeq
Replacing $\beta(n,1)$ in equation  \refbr{total.amd} by \refbr{beta.1.zal}
we will get:
\beq
T(n,p)=\beta(n,p)\cdot
T(n,p)+\frac{T(n,1)}{p}-\frac{\beta(n,p)\cdot T(n,p)}{p}
\eeq
Now, multiplying both sides by  $p$ and moving all components with  $T(n,p)$
to the left hand side we receive:
\beq
p \cdot T(n,p)-p\cdot\beta(n,p)\cdot
T(n,p)+\beta(n,p)\cdot T(n,p)=T(n,1)
\label{GB-przedost}
\eeq
We will get the value of speedup $S(n,p)$ obtained owing to the parallelization
dividing both sides of the equation \refbr{GB-przedost} by
$T(n,p)$. So, it will be equal:
\beq
S(n,p)=\frac{T(n,1)}{T(n,p)}=p-(p-1)\cdot\beta(n,p)
\label{prawo.GB}
\eeq
In this way we received nothing but Gustafson-Barsis law \refbr{GustBar.oryg}.

What concerns the better speedup in Gustafson's experiments with the growing
size of the problem (and the number of processors which was linked there) we may explain it
in the following way. Gustafson assumed, that the time spent in the serial part
("for vector startup, program loading, serial bottlenecks, I/O operations") do not depend on the problem size, that is
\beq
T_s(n) = const. = T_s = \beta(n,1)\cdot T(n,1)=\beta_s \cdot T(1,1),\;\; \forall n
\eeq
while the total time of the execution of the parallel part on the sequantial
machine was proportional to the problem size
$n$.
In this way the serial factor on the sequential machine $\beta(n,1)$ was equal
 \beq
\beta(n,1) =
\frac{\beta_s \cdot T(1,1)}{\beta_s \cdot T(1,1) +n\cdot(1-\beta_s)\cdot T(1,1)} =
\frac{\beta_s}{\beta_s+n\cdot(1-\beta_s)} = 
\frac{1}{1
+n\cdot (\frac{1}{\beta_s} -1)}
\eeq
A similar situation would be when the time $T_s(n)$ is proportional to the problem size $n$ (e.g.~$n \cdot
\beta_s$), but the time spent  in the parallel part is proportional to ~$n^2$ \hbox{(e.g.~$n^2 \cdot (1-\beta_s)$)}.
In such cases
\beq
\beta(n,1) \goesto_{n \goesto \infty} 0
\eeq
what means, taking into account \refbr{pr.Amd}, that
\beq
S(n,p) \goesto_{n \goesto \infty} p
\eeq
In other words, also from Amdahl's law we may conclude, that the bigger the
size of the problem,  the closer the speedup to the number
of processors.

\section{Conclusions}

In the paper it is shown, that the Gustafson-Barsis law can be directly derived
from the Amdahl's law, without strange assumptions as normalizng to one the
time of execution of the program on the sequential machine. Moreover, the
speedups approaching the number of processors observed in the experiments
described in the Gustafson's paper can be concluded from the Amdahl's law,
when we take into account as the arguments of the serial factor the size
of the problem and the number of processors.



\section*{Acknowledgments}

This research was supported by the Polish Ministry of Science and Higher Education
under grant N N514 416934 (years 2008-2010).

\end{document}